\documentclass[aps,pre,
twocolumn,
amssymb,amsmath,
showpacs,showkeys,
superscriptaddress,
]{revtex4}
\usepackage[final]{graphicx}
\usepackage{dcolumn}
\begin{document}
\title{Recursive Schr\" odinger Equation Approach to Faster Converging Path Integrals}
\author{Antun Bala\v{z}}\email[E-mail: ]{antun@phy.bg.ac.yu}
\author{Aleksandar Bogojevi\'c}
\author{Ivana Vidanovi\'c}
\affiliation{Scientific Computing Laboratory, Institute of Physics Belgrade, Pregrevica 118, 11080 Belgrade, Serbia}
\homepage[Home page: ]{http://scl.phy.bg.ac.yu/}
\author{Axel Pelster}
\affiliation{Fachbereich Physik, Universit\" at Duisburg-Essen, Lotharstra\ss e 1, 47048 Duisburg, Germany}
\affiliation{Institut f\"ur Theoretische Physik, Freie Universit\"at Berlin, Arnimallee 14, 14195 Berlin, Germany}

\begin{abstract}
By recursively solving the underlying Schr\" odinger equation,
we set up an efficient systematic approach for deriving analytic expressions for discretized effective actions.
With this we obtain discrete short-time propagators for both one and many particles in arbitrary dimension to
orders which have not been accessible before. They
can be used to substantially speed up
numerical Monte Carlo calculations of path integrals, as well as for setting up a new analytical approximation scheme for
energy spectra, density of states, and other statistical properties of quantum systems.
\end{abstract}
\preprint{SCL preprint}
\maketitle

\section{Introduction}

The central object in the path-integral formulation of quantum statistics
is the (Euclidean) transition amplitude $A(a,b;t)=\langle b|e^{-t\hat H}|a\rangle$
\cite{feynman,feynmanhibbs,feynmanstat,kleinertbook}. The starting point in
setting up this formalism is the completeness relation
\begin{eqnarray}
\label{complete}
A(a,b;t)&=&\int dq_1\cdots \int dq_{N-1}\nonumber \\
&& \times \,A(a,q_1;\varepsilon)\cdots A(q_{N-1},b;\varepsilon)\, ,
\end{eqnarray}
where $\varepsilon=t/N$ denotes the time-slice width. To leading order in $\varepsilon$  the short-time
transition amplitude reads in natural units
\begin{eqnarray}
\label{lowest}
\lefteqn{
A(q_n,q_{n+1};\varepsilon)=}\nonumber\\
&&=\frac{1}{(2\pi\varepsilon)^{d/2}}\exp \left[-\frac{(q_{n+1}-q_n)^2}{2\varepsilon}-\varepsilon V(x_n)\right]\, ,
\end{eqnarray}
where the potential $V$ is evaluated
at the mid-point coordinate $x_n=(q_n+q_{n+1})/2$.
Substituting (\ref{lowest}) in the completeness relation (\ref{complete}), the deviation of the
obtained discrete amplitude from the continuum result turns out to be of the order $O(\varepsilon)$.
This slow convergence to the continuum is the major reason for the low efficiency of the ubiquitous
Path Integral Monte Carlo simulations \cite{ceperley}, especially in numeric studies of Bose-Einstein condensation
phenomena \cite{pilati2006,zhang,pilati2008}, quantum phase transitions and phase diagrams at low temperatures  \cite{vitali,kim}.
Thus, in order to accelerate numerical calculations for statistical properties of quantum systems, it is indispensable to
develop more efficient algorithms. The existence of such algorithms has been established recently \cite{predescu}.

To this end we worked out and numerically verified in a series of recent
papers \cite{prl-speedup,prb-speedup,pla-euler,pre-ideal}
an efficient analytical procedure for
improving the convergence of path integrals for single-particle
transition amplitudes to the order $O(\varepsilon^p)$ for arbitrary values of $p$.
This was achieved by studying how
discretizations of different coarseness are related to a hierarchy of effective discrete-time
actions which improve the convergence in a systematic way.
In Ref.~\cite{pla-manybody} we presented an equivalent approach which is based
on a direct path-integral calculation of intermediate time amplitudes to the order $O(\varepsilon^p)$.
The inherent simplicity
of these direct calculations made it possible to extend the procedure to general many-body theories
in arbitrary dimension and to obtain
explicit results for the effective actions up to level $p=5$. It turned out that
increasing $p$ leads to an exponential rise in
complexity of the effective actions which, ultimately, limits the maximal value of $p$ one can practically
work with. These
limitations of existing approaches, in particular in the case of many-body theories,
are still below the calculational barrier stemming from this rise in
complexity. This is a strong indication that new
and more efficient calculational schemes must exist which should considerably improve
the convergence of path integrals for general many-body theories. The availability of analytic
expressions for higher $p$ effective
actions is essential for numerical calculations of path integrals with high precision. Obtaining the information on energy spectra
is just one important example of calculations that require high-precision numerical results. Furthermore, since the structure of
higher order terms of effective actions is governed by the quantum dynamics of the system, it can be
used for extracting analytical information about the system properties.

As is well known, in concrete calculations it is always easier to solve the underlying Schr\"odinger equation than
to directly evaluate the corresponding path integral. For instance, in the case of particular potentials
the Schr\" odinger equation approach allows  an efficient recursive scheme
to calculate perturbation series up to very high orders \cite{S1,S2,S3,S4,S5}. With this in mind,  we
develop in the present paper a new and more efficient recursive approach for deriving
the short-time transition amplitude from the underlying
Schr\" odinger equation. To this end we proceed as follows.
Section 2 presents the instructive case of a single one-dimensional particle
moving in a general potential. The maximal level obtained by the new method is $p=35$, and thus compares favorably with
the best result $p=9$ of previous approaches. In Section 3 we focus on the restricted problem of evaluating
the velocity independent part of the discrete-time effective potential. We derive the differential
equations for the velocity independent effective potential of a single one-dimensional particle and
solve them for the case of a general potential up to level $p=37$. Section 4 extends the results of Section 2
to the case of a general many-body theory in $d$ dimensions. We derive the differential equations for the general
many-body effective potential and solve them up to level $p=10$. Both the equations and their
solution are presented in Section 5 in a diagrammatic form which is similar to the
recursive graphical construction of Feynman diagrams worked out in the series
\cite{C1,C2,C3,C4,C5,C6}.
With this we illustrate the inherent combinatoric nature of determining the discrete-time effective
potential. The growing number of diagrams with level $p$ leads to an increasing complexity of the
expression for the discrete-time effective potential. We end by commenting on how this growth
in complexity limits maximal attainable values of $p$. Throughout the paper we describe some important envisaged applications
of the derived discretized effective actions for calculating statistical properties of relevant quantum systems. These applications
include not only numerical calculations but also a new analytical approximation scheme which was made possible through the
availability of high level effective actions.
\section{One particle in One Dimension}
We start with calculating the transition amplitude $A(q,q';\varepsilon)$ for one particle in one dimension.
It obeys the symmetry
\begin{eqnarray}
\label{SYM}
A(q,q';\varepsilon)=A(q',q;\varepsilon)
\end{eqnarray}
and satisfies the time-dependent Schr\" odinger equations
\begin{eqnarray}
\left[\frac{\partial}{\partial \varepsilon}-\frac{1}{2}\,\frac{\partial^2}{\partial q^2}+V(q)\right]
A(q,q';\varepsilon)&=&0\, , \\
\left[\frac{\partial}{\partial \varepsilon}-\frac{1}{2}\,\frac{\partial^2}{\partial q'^2}+V(q')\right]
A(q,q';\varepsilon)&=&0
\end{eqnarray}
with the initial condition
\begin{equation}
\label{initial}
A(q,q';0)=\delta(q-q') \, .
\end{equation}
In terms of the deviation $\bar x=(q'-q)/2$ and
the mid-point coordinate $x=(q+q')/2$, both equations are rewritten according to
\begin{eqnarray}
\label{symmetric}
\left[\frac{\partial}{\partial \varepsilon}-\frac{1}{8}\,\partial^2-\frac{1}{8}\,\bar\partial^2
+\frac{1}{2}\, (V_++V_-)\right]A&=&0\, , \\
\label{antisymmetric}
\left[-\, \partial\bar\partial +2\, (V_+-V_-)\right]A&=&0\, ,
\end{eqnarray}
where we have introduced  $V_\pm=V(x\pm\bar x)$ as an abbreviation. Their solution may be written in the form
\begin{equation}
\label{exact}
A=\frac{1}{\sqrt{2\pi \varepsilon}}\exp \left[-\frac{2}{\varepsilon}\,\bar x^2-\varepsilon W(x,\bar x;\varepsilon)
\right]\, ,
\end{equation}
where the effective potential $W(x,\bar x;\varepsilon)$ is an even function of $\bar x$ due to
the symmetry (\ref{SYM}) of the Euclidean transition amplitude. Note that
Eq.~(\ref{lowest}) represents an approximation to the exact form (\ref{exact}) up to order $O(\varepsilon)$.
Substituting (\ref{exact}) in (\ref{symmetric}) and (\ref{antisymmetric}) yields
\begin{eqnarray}
\label{first}
& &\hspace*{-1.0cm}
W+\bar x\,\frac{\partial W}{\partial\bar x}+\varepsilon\,\frac{\partial W}{\partial \varepsilon}
-\frac{1}{8}\,\varepsilon\,\partial^2 W-
\frac{1}{8}\,\varepsilon\,\bar\partial^2 W\nonumber\\
& &\hspace*{-1.0cm}
\quad\ +\,\frac{1}{8}\,\varepsilon^2\,(\partial W)^2
+\frac{1}{8}\,\varepsilon^2\,(\bar\partial W)^2= \frac{1}{2}\, (V_++V_-)\, ,\\
\label{second}
& & \hspace*{-1.0cm}
\bar x\,\partial W-\frac{1}{4}\,\varepsilon\,\partial\bar\partial W
+\frac{1}{4}\,\varepsilon^2\,\partial W\bar\partial W = \frac{1}{2}\,(V_+-V_-)\, .
\end{eqnarray}
Both partial differential equations determine the effective potential $W(x,\bar x;\varepsilon)$
and thus the transition  amplitude $A(q,q';\varepsilon)$.
The initial condition (\ref{initial})
implies that $W$ is regular in the vicinity of $\varepsilon=0$, i.e. it may be expanded in a power
series in $\varepsilon$. We are interested in using $W$ to systematically speed up the convergence of discrete amplitudes
with $N$ time slices to the continuum limit. This is done by evaluating $W$ to higher powers in $\varepsilon$.
According to Eq.~(\ref{lowest}) the
dominant term for the short-time propagation is
the diffusion relation $\bar x^2\propto \varepsilon$.
Therefore, we expand $W$ in a double power series in both $\varepsilon$ and $\bar x^2$:
\begin{equation}
\label{ansatz}
W(x,\bar x;\varepsilon)=\sum_{m=0}^{\infty}\sum_{k=0}^{m}c_{m,k}(x)\,\varepsilon^{m-k}\bar x^{2k}\ .
\end{equation}
Restricting the sum over $m$ from 0 to $p-1$ leads to a discrete amplitude that converges to the continuum result as
$\varepsilon^p$, i.e. as $1/N^p$. For later convenience we define all coefficients $c_{m,k}$, which are
not explicitly used in Eq.~(\ref{ansatz}), to be zero, i.e. we set $c_{m, k}=0$ whenever the condition
$m\geq k\geq 0$ is not satisfied.

\begin{figure}[t]
\centering
\includegraphics[width=0.3 \textwidth]{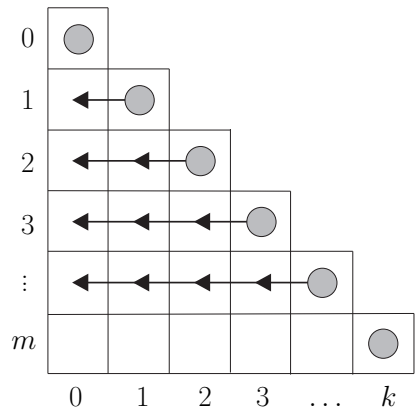}
\caption{(Color online) Order in which the coefficients $c_{m,k}$ are calculated. Diagonal ones follow
from Eq.~(\ref{diagonal}), off-diagonal ones from the recursion relation (\ref{1drec}).}
\label{order}
\end{figure}

Substituting the expansion (\ref{ansatz})
into the partial differential equations (\ref{first}) and (\ref{second}) leads to two equivalent recursion relations.
The second recursion relation turns out to be more difficult to solve to higher orders as it directly determines
not the coefficients $c_{m,k}(x)$ but their first derivatives. For this reason we
restrict ourselves in the remainder of this section to the recursion
relation following from Eq.~(\ref{first}). The diagonal coefficients are given by
\begin{equation}
\label{diagonal}
c_{m,m}=\frac{V^{(2m)}}{(2m+1)!}\, ,
\end{equation}
while off-diagonal coefficients satisfy the recursion relation
\begin{eqnarray}
&&\hspace*{-1cm}
8(m+k+1)\, c_{m,k}=(2k+2) (2 k+1)\, c_{m,k+1} \nonumber\\
&&\hspace*{-0.5cm}
\,\, +c_{m-1,k}''-\sum_{l=0}^{m-2}\, \sum_{r} c_{l,r}'\,c_{m-l-2,k-r}'\nonumber\\
&&\hspace*{-0.5cm}
\,\, -\sum_{l=1}^{m-2}\,\sum_{r}2\,r(2k-2r+2)\,c_{l,r}\,c_{m-l-1,k-r+1}\, ,
\label{1drec}
\end{eqnarray}
where the sum over $r$ goes from ${\rm max}\{0, k-m+l+2\}$ to
${\rm min}\{k,l\}$ in accordance with the restriction that $c_{m, k}=0$ whenever the condition
$m\geq k\geq 0$ is not satisfied.
For a given value of $m$, the coefficients $c_{m,k}$ for $k=0,1,\ldots,m$ are determined as follows.
The diagonal coefficient $c_{m,m}$ is directly given by (\ref{diagonal}), whereas the off-diagonal
coefficients $c_{m,k}$ follow recursively from evaluating (\ref{1drec}) for $k=m-1,\ldots,1,0$. This
recursive solution method is schematically depicted in Fig.~\ref{order}. Let us illustrate this procedure
for the lowest levels. For $p=1$ we immediately obtain from (\ref{diagonal})
\begin{equation}
\label{c00}
c_{0,0} = V \, .
\end{equation}
For $p=2$ we have to first determine $c_{1,1}$ from (\ref{diagonal}), yielding
\begin{equation}
\label{c11}
c_{1,1} = \frac{V''}{6} \, .
\end{equation}
Then recursion relation (\ref{1drec}) states that $c_{1,0}$ follows from the previously determined coefficients
according to
\begin{equation}
\label{c10}
c_{1,0} = \frac{1}{16}c''_{0,0} + \frac{1}{8}c_{1,1}\, .
\end{equation}
From (\ref{c00})--(\ref{c10}) we then read off the result
\begin{equation}
c_{1,0} =\frac{V''}{12} \, .
\end{equation}
Similarly, we find for $p=3$
\begin{eqnarray}
&&\hspace*{-1cm}
c_{2,2} = \frac{V^{(4)}}{120}\, ,\\
&&\hspace*{-1cm}
c_{2,1} = \frac{1}{32}\,c''_{1,1}+\frac{3}{8}\,c_{2,2}=\frac{V^{(4)}}{120}\, ,\\
&&\hspace*{-1cm}
c_{2,0} = \frac{1}{24}\,c''_{1,0}+\frac{1}{12}\,c_{2,1}-\frac{1}{24}\,(c'_{0,0})^2=\frac{V^{(4)}}{240}-\frac{V'^2}{24}\, .
\end{eqnarray}
The outlined procedure continues in the same way for higher levels $p$. We have automatized this procedure and
implemented it using the Mathematica 6.0 package \cite{mathematica} for symbolic calculus. Using this we determined
the effective action for a one-dimensional particle in a general potential
up to the level $p=35$.
Although the effective actions grow in complexity with level $p$, the Schr\" odinger equation method for
calculating the discrete-time effective actions turns out to be
extremely efficient. The ultimate value of $p=35$ far surpasses the previously obtained best result of $p=9$,
and is only limited by the sheer size of the expression for the effective
action of a general theory at such a high level.

The whole technique can be pushed much further when working on specific potential classes. For example, for a
particle moving in a quartic potential we have obtained effective actions up to $p=140$. Similar levels have been
achieved for higher-order polynomial potentials. The increase in level and the decrease in size of the expressions for
the effective actions originate from functional relations between the potential and its
derivatives. These relations are particularly simple in the case of polynomial interactions where all the derivatives
of the potential above a certain degree vanish. However,
the benefits of working within a specific class of potentials are
not only limited to polynomial interactions. For example, the functional relations for the modified
P\" oschl-Teller potential have allowed us to obtain effective actions up to level $p=40$ \cite{speedup}.

As already stated, the principal rationale behind constructing high-level effective actions is to use them for
speeding up Monte Carlo calculations. However, having obtained explicit expressions for effective actions to such
high levels, it now becomes possible to use them extensively in both numerical and analytical calculations. In
particular, having obtained an extremely precise knowledge of the short-time propagation of a system,
it is possible to use standard resummation techniques such as the Pad\'e and the Borel method to extract useful information
about its behavior for long propagation times.

The derived effective actions can also be applied to systematically improve the Numerical Matrix Diagonalization (NMD)
method \cite{pollockceperley, sethia1990, sethia1999} for calculating energy eigenvalues and eigenstates. Note that the propagation
time $t$ used in the NMD method is just a parameter which is chosen in such a way that it minimizes the error associated with the
calculated energy eigenvalues. Therefore, it is always possible to select this parameter to be small, so that the obtained expansion
of the ideal effective action can be used to substantially improve NMD calculations. Furthermore, in this case we can even use
analytic $N=1$ approximation for the path integral. In this approximation there are no integrals to perform in Eq. (\ref{complete})
and the amplitude is directly given by the analytic expression (\ref{exact}). Using such extremely rough discretizations is only
possible if one has determined the ideal effective action to very high orders $p$. In effect, one compensates without loss of
precision the increase of discretization coarseness with the input of new analytical information concerning the propagation time
which is contained in the effective action. In this way, without any integration or resummation techniques, we can calculate large
number of highly accurate energy eigenvalues, avoiding the usually needed limit $t\to\infty$ which is difficult to approach. The
large number of precise energy eigenvalues obtained by using this improved NMD method allows for calculating amplitudes for longer
propagation times using the spectral decomposition. This also makes it possible to calculate partition functions as well as global,
local, and bilocal densities of states or other relevant statistical quantities with high accuracy.
\section{Velocity Independent Part of Effective Potential}
The velocity independent part of the effective potential $W_0(x;\varepsilon)\equiv W (x,0;\varepsilon)$
determines diagonal transition amplitudes.
Although it does not contain all the information which is needed for constructing
the effective action, it is of interest for determining physical quantities such
as the particle density and the energy spectra.
This object is also much simpler than the full $W$. In addition, the relation between $V$
and $W_0$ allows us to visualize the effects of quantization and discretization for a given potential.
In this section we derive
the differential equation for $W_0$ for a single particle moving in one dimension.

Both Eqs.~(\ref{symmetric}) and (\ref{antisymmetric}) contain derivatives with respect to $\bar x$, so it is impossible
to set $\bar x=0$ and obtain equations for $W_0$. However, differentiating Eq.~(\ref{antisymmetric})
with respect to $\bar x$ we get
\begin{equation}
\partial\bar\partial^2 A=4\,\partial V A+\ldots\ ,
\end{equation}
where the dots denote terms which vanish when $\bar x\to 0$.
Thus, differentiating Eq.~(\ref{symmetric}) with
respect to $x$ and substituting the above result we obtain an equation which does not contain derivatives
with respect to $\bar x$.
Finally, setting $\bar x=0$ we find the differential equation for diagonal transition amplitudes
\begin{equation}
\left(\partial\frac{\partial}{\partial\varepsilon}-\frac{1}{8}\,\partial^3+\frac{1}{2}\,\partial V
+V\partial\right)
A(x,x;\varepsilon)=0\ .
\end{equation}
Substituting $A(x,x;\varepsilon)=(2\pi\varepsilon)^{-1/2}\exp (-\varepsilon W_0)$ then yields the equation for $W_0$:
\begin{eqnarray}
&&\hspace*{-0.5cm}
\partial W_0+\frac{\varepsilon}{4}\left(8\,\partial\frac{\partial W_0}{\partial\varepsilon}
-8W_0\,\partial W_0+8V\partial W_0-\partial^3 W_0\right)\\
& &\hspace*{-0.3cm}
-\frac{\varepsilon^2}{4}\left(8\,\partial W_0\frac{\partial W_0}{\partial\varepsilon}
-3\partial W_0\partial^2 W_0\right)
-\frac{\varepsilon^3}{4}\,(\partial W_0)^3=\partial V\, .\nonumber
\label{W0}
\end{eqnarray}
This is solved in the form of the power series
\begin{equation}
W_0(x;\varepsilon)=\sum_{m=0}^\infty c_{m,0}(x)\varepsilon^m\, .
\end{equation}
Inserting this into the differential equation (\ref{W0})
determines the coefficients $c_{m,0}$  through the simple recursion relation
\begin{eqnarray}
&& \hspace*{-0.6cm}
(2m+1)\, c_{m,0}'= \frac{1}{4}\,c_{m-1,0}'''-2\, V c_{m-1,0}'
\\
&& \hspace*{-0.6cm}+2\sum_{k=0}^{m-1} c_{k,0}'\, c_{m-k-1,0}
+ 2\sum_{k=1}^{m-1} k\, c_{k,0}\, c_{m-k-1,0}'
\nonumber\\
&&\hspace*{-0.6cm}
-\frac{3}{4}\sum_{k=0}^{m-2} c_{k,0}'\, c_{m-k-2,0}''
+\frac{1}{4}\sum_{k=0}^{m-3}\,\sum_{l=0}^{m-k-3} c_{k,0}'\, c_{l,0}'\, c_{m-k-l-3,0}'\, .\nonumber
\end{eqnarray}
With this we have evaluated the velocity independent part of the effective potential up to level $p=37$
for a particle moving in a generic potential $V(x)$. As before, for specific potential classes one can go to much
higher levels.
\section{Many-Body Systems}
Now we extend the calculations of Section 2 to the case of a general non-relativistic theory of $M$ particles in $d$
dimensions. The derivation of the equations for $W$ proceeds completely parallel to the case of one particle in one
dimension. The Schr\" odinger equations now have the form
\begin{eqnarray}
\label{SCH1}
\left[\frac{\partial}{\partial \varepsilon}-\frac{1}{2}\,\sum_{i=1}^M\triangle_i+V(q)\right]
A(q,q';\varepsilon)&=&0\, , \\
\label{SCH2}
\left[\frac{\partial}{\partial \varepsilon}-\frac{1}{2}\,\sum_{i=1}^M\triangle'_i+V(q')\right]
A(q,q';\varepsilon)&=&0\, ,
\end{eqnarray}
where $\triangle_i$ and $\triangle'_i$ stand for $d$-dimensional Laplacians over initial and final coordinates of
particle $i$, while $q$ and $q'$ are $d \times M$
dimensional vectors representing positions of all particles at the initial
and final moment. Furthermore, the potential $V$ contains both the external potential and the respective
interaction potentials between two and more particles.
After substituting the $dM$-dimensional generalization of the expression for the transition amplitude (\ref{exact}) into
the Schr\" odinger equations (\ref{SCH1}) and (\ref{SCH2}),
we find the analogues of Eqs.~(\ref{first}) and (\ref{second}) for the effective potential $W$
\begin{eqnarray}
&&\hspace*{-1.0cm}
W+\bar x\cdot\bar\partial\,W+\varepsilon\frac{\partial W}{\partial \varepsilon}
-\frac{1}{8}\,\varepsilon\partial^2 W-\frac{1}{8}\,\varepsilon\bar\partial^2 W\nonumber\\
&&\hspace*{-0.5cm}+\frac{1}{8}\,\varepsilon^2(\partial W)^2
+\frac{1}{8}\,\varepsilon^2(\bar\partial W)^2=\frac{1}{2}\, (V_++V_-)\, ,\label{E1}\\
&&\hspace*{-1.0cm}\bar x\cdot\partial\, W-\frac{\varepsilon}{4}\,\partial\cdot\bar\partial\, W
+\frac{\varepsilon^2}{4} (\partial W)\cdot(\bar\partial W)
\nonumber\\&&\hspace*{-0.5cm}
=\frac{1}{2}\, (V_+-V_-)\, .
\end{eqnarray}
Here we have used the definition
$A\cdot B=A_iB_i$, where $i=1,2,\ldots, Md$ and repeated indices are summed over. Either of
the above two equations for $W$ can be used to determine the appropriate short-time expansion as a double Taylor
series in powers of $\varepsilon$ and even powers of $\bar x$:
\begin{equation}
\label{EXP}
W(x,\bar x;\varepsilon) = \sum_{m=0}^{\infty} \, \sum_{k=0}^{m}\varepsilon^{m-k}\,W_{m,k}(x,\bar x)\ ,
\end{equation}
where $W_{m,k}(x,\bar x)= \bar x_{i_1} \bar x_{i_2} \cdots \bar x_{i_{2k}}c_{m,k}^{i_1,\ldots, i_{2k}}(x)$.
It turns out to be
advantageous to use recursion relations for the fully contracted quantities $W_{m,k}$ rather than the
respective coefficients
$c_{m,k}^{i_1,\ldots, i_{2k}}$. In this way we avoid the computationally expensive symmetrization over
all indices $i_1,\ldots, i_{2k}$. Again it is easier to work with the first of the two
equations for $W$. Substituting (\ref{EXP}) into (\ref{E1}) directly yields the diagonal coefficients
\begin{equation}
\label{DIA}
W_{m,m}=\frac{1}{(2m+1)!}(\bar x\cdot\partial)^{2m}\,V\, .
\end{equation}
The off-diagonal coefficients satisfy the recursion relation which represents a generalization of
Eq.~(\ref{1drec}):
\begin{eqnarray}
\label{recursion}
&&\hspace*{-1.0cm}
8\, (m+k+1)\,W_{m,k}=\partial^2 W_{m-1,k}+\bar\partial^2 W_{m,k+1}\nonumber\\
&&\hspace*{-0.5cm}-\sum_{l=0}^{m-2}\,\sum_{r}(\partial W_{l,r})\cdot(\partial W_{m-l-2,k-r})\nonumber\\
&&\hspace*{-0.5cm}-\sum_{l=1}^{m-2}\,\sum_{r}(\bar\partial W_{l,r})\cdot(\bar\partial W_{m-l-1,k-r+1})\, .
\label{REC}
\end{eqnarray}
As before, the sum over $r$ goes from ${\rm max}\{0, k-m+l+2\}$ to ${\rm min}\{k,l\}$.
The above recursion disentangles, in complete analogy with the previously outlined case of one particle in one dimension.
To illustrate this we write down and solve the equations up to level $p=4$:
\begin{eqnarray}
W_{0,0}&=&V\, ,\nonumber\\
W_{1,1}&=&\frac{1}{6}\,(\bar x\cdot\partial)^2 V\, ,\nonumber\\
W_{1,0}&=&\frac{1}{16}\,\partial^2 W_{0,0}+\frac{1}{16}\,\bar\partial^2 W_{1,1}=\frac{1}{12}\partial^2 V\, ,\nonumber\\
W_{2,2}&=&\frac{1}{120}\,(\bar x\cdot\partial)^4 V\, ,\nonumber\\
W_{2,1}&=&\frac{1}{32}\,\partial^2 W_{1,1}+\frac{1}{32}
\bar\partial^2 W_{2,2}=\frac{1}{120}(\bar x\cdot\partial)^2 \partial^2 V\, ,\nonumber
\end{eqnarray}
\begin{eqnarray}
W_{2,0}&=&\frac{1}{24}\,\partial^2 W_{1,0}+\frac{1}{24}\,\bar\partial^2 W_{2,1}
-\frac{1}{24}\,(\partial W_{0,0})^2\nonumber\\
&=&\frac{1}{240}\,\partial^4 V-\frac{1}{24}\,(\partial V)\cdot(\partial V)\, ,\nonumber\\
\label{solution}
W_{3,3}&=&\frac{1}{5040}\,(\bar x\cdot\partial)^6 V\, ,\nonumber\\
W_{3,2}&=&\frac{1}{48}\,\partial^2 W_{2,2}+\frac{1}{48}
\,\bar\partial^2 W_{3,3}=\frac{1}{3360}(\bar x\cdot\partial)^4\partial^2 V\, ,\nonumber\\
W_{3,1}&=&\frac{1}{40}\,\partial^2 W_{2,1}+\frac{1}{40}\,\bar\partial^2 W_{3,2}
\nonumber\\&&
-\frac{1}{20}(\partial W_{0,0})\cdot(\partial W_{1,1})
-\frac{1}{40}(\bar\partial W_{1,1})^2
\nonumber\\&=&
\frac{1}{3360}\,(\bar x\cdot\partial)^2\partial^4 V-\frac{1}{120}
\,\partial_i V(\bar x\cdot\partial)^2\partial_i V\nonumber\\
&&-\,\frac{1}{360}\,(\bar x\cdot\partial)\partial_i V\, (\bar x\cdot\partial)\partial_i V\, ,\nonumber
\end{eqnarray}
\begin{eqnarray}
W_{3,0}&=&\frac{1}{32}\,\partial^2 W_{2,0}+\frac{1}{32}\,\bar\partial^2 W_{3,1}
-\frac{1}{16}\,(\partial W_{0,0})\cdot(\partial W_{1,0})\nonumber\\
&=&\frac{1}{6720}\,\partial^6 V-\frac{1}{120}\,\partial_i V \partial^2 \partial_i V
\nonumber\\&&
-\,\frac{1}{360}\,\partial_i\partial_j V \partial_i\partial_j V\, .
\end{eqnarray}
\section{Graphical Representation}
The above equations and their solutions can be cast
in a diagrammatic form which is similar to the
recursive graphical construction of Feynman diagrams worked out in the series
\cite{C1,C2,C3,C4,C5,C6}. The effective potential $W$ represents the sum of all connected vacuum diagrams
of the underlying theory with the following Feynman rules.
The propagator is represented by the
Kronecker delta
\begin{equation}
\end{equation}
\begin{center}
\vspace*{-0.8cm}
\includegraphics[width=3.5cm]{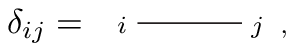}
\end{center}
the $l$-point vertex is the $l$-th derivative of the potential
\vspace*{0.3cm}
\begin{equation}
\end{equation}
\begin{center}
\vspace*{-0.8cm}
\includegraphics[width=4.5cm]{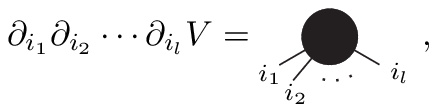}
\end{center}
and the even number of external sources stand for the discrete velocities $\bar x$
\begin{equation}
\end{equation}
\begin{center}
\vspace*{-0.8cm}
\includegraphics[width=2.8cm]{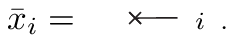}
\end{center}
A simple dimensional analysis determines those diagrams which contribute to a
given coefficient $W_{m,k}$. The discrete-time effective potential $W$ is the generating functional
of connected diagrams since it appears in the exponent. The diagrammatic notation makes it explicit that the
short-time expansion of the effective discrete potential is a purely combinatoric problem in which all the information
is contained in the symmetry factors multiplying individual diagrams. Thus, Eq.~(\ref{recursion}) represents the
Schwinger-Dyson equation of the underlying theory. As such it is the simplest way for actually determining the symmetry
factors.

The Schwinger-Dyson equation is now cast in a diagrammatic form. To this end we begin with
introducing general diagrams for $W_{m, k}$
\vspace*{0.4cm}
\begin{equation}
\end{equation}
\begin{center}
\vspace*{-1.4cm}
\includegraphics[width=4.0cm]{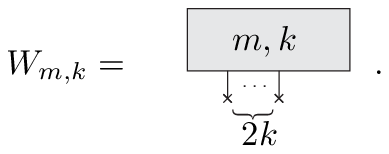}
\end{center}
According to (\ref{DIA})
diagonal terms $W_{m,m}$ are directly given in terms of vertices which are contracted with even numbers of external
sources:
\vspace*{0.2cm}
\begin{equation}
\end{equation}
\begin{center}
\vspace*{-1.2cm}
\hspace*{-0.5cm}
\includegraphics[width=7.4cm]{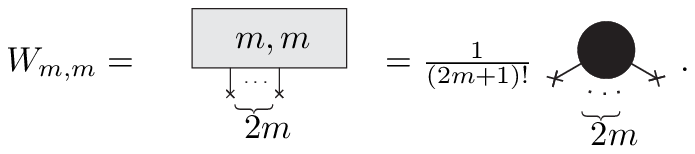}
\vspace*{0.2cm}
\end{center}

\begin{figure}[!b]
\centering
\includegraphics[width=8.4cm]{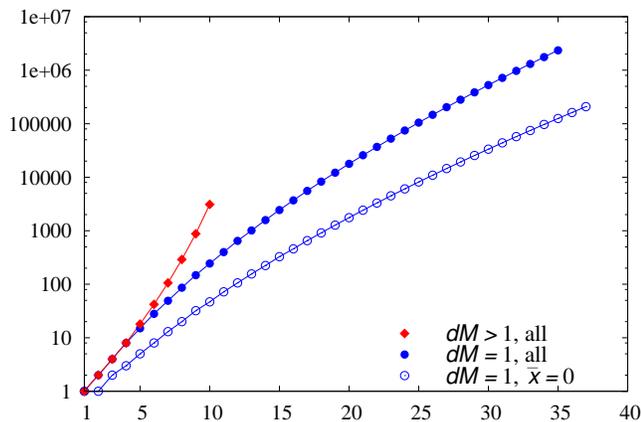}
\caption{(Color online) The number of diagrams contributing to a given level illustrates the rise in complexity with
increasing $p$. Systems fall into one of two classes of complexity: the more
complex ones with $dM>1$ (top curve) and the less complex ones with $dM=1$ (middle curve). In addition, the
 bottom curve gives the number of velocity independent diagrams for $dM=1$.}
\label{terms}
\end{figure}

\noindent
The off-diagonal recursion relation (\ref{REC})
contains a differentiation of diagrams with respect to the mid-point coordinate $x$ and the
discrete velocity $\bar x$. These operations act as follows:
\vspace*{0.4cm}
\begin{equation}
\end{equation}
\begin{center}
\vspace*{-1.0cm}
\hspace*{-0.6cm}
\includegraphics[width=6.3cm]{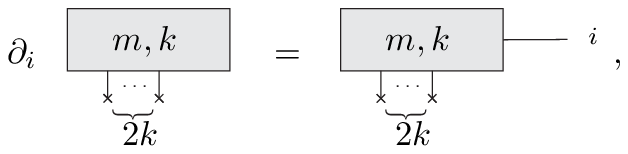}
\end{center}
\vspace*{-0.4cm}
\begin{equation}
\end{equation}
\begin{center}
\vspace*{-1.5cm}
\includegraphics[width=6.6cm]{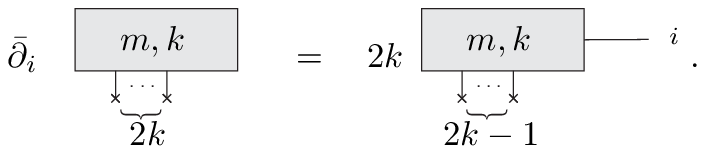}
\end{center}
Putting all those elements together we find the graphical representation of the Schwinger-Dyson equation:
\begin{widetext}
\begin{equation}
\end{equation}
\begin{center}
\vspace*{-1.1cm}
\includegraphics[width=17cm]{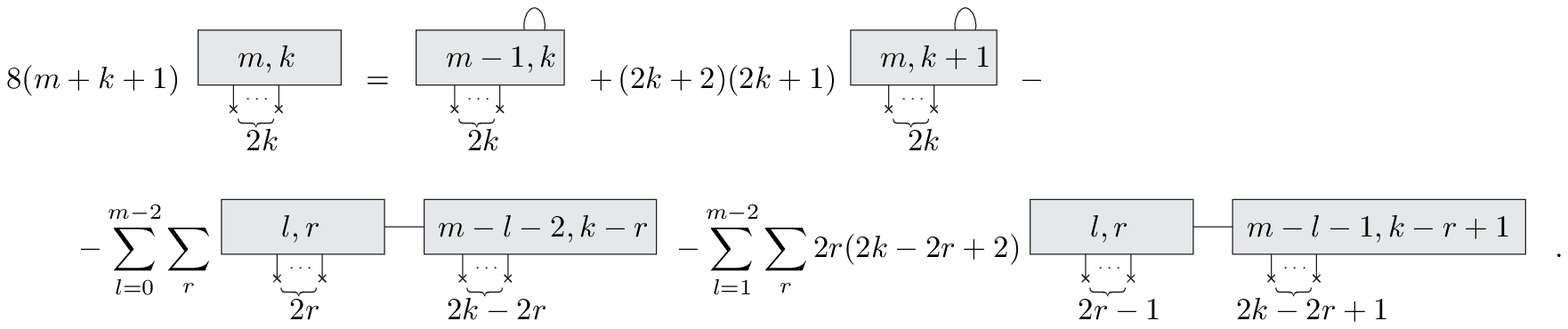}
\end{center}
\vspace*{0.2cm}
\end{widetext}
The sum over $r$ has the range as defined after Eq.~(\ref{recursion}).
From this we read off via complete induction that, indeed, all vacuum diagrams contributing to the
effective potential are connected.
In this diagrammatic notation, the previously obtained solutions for the discrete-time
effective potential of a general many-body theory up to level $p=4$ read:
\begin{widetext}
\begin{flushleft}
\includegraphics[height=6.3cm]{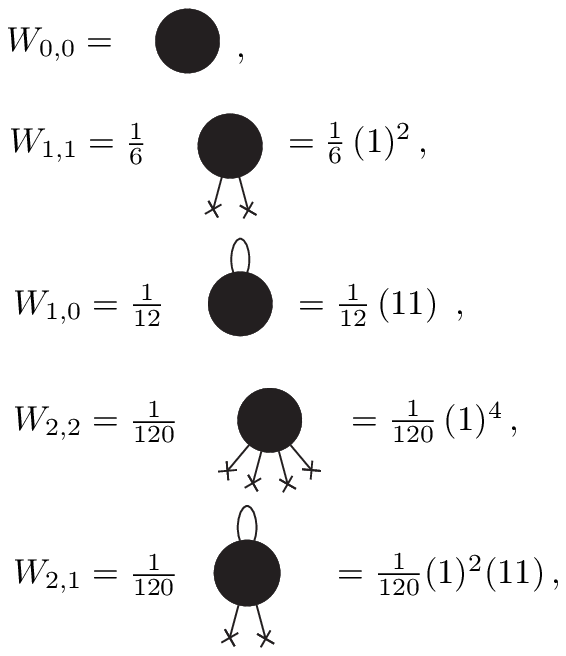}\hspace*{3.7cm}
\includegraphics[height=5.8cm]{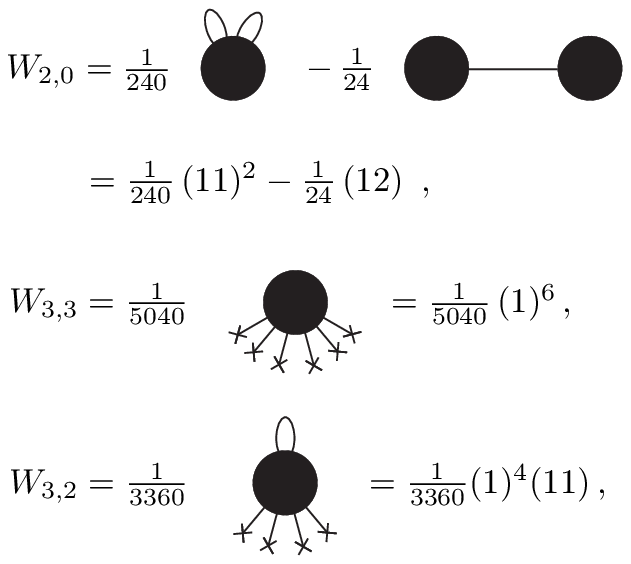}\\
\vspace*{0.4cm}
\includegraphics[width=16.4cm]{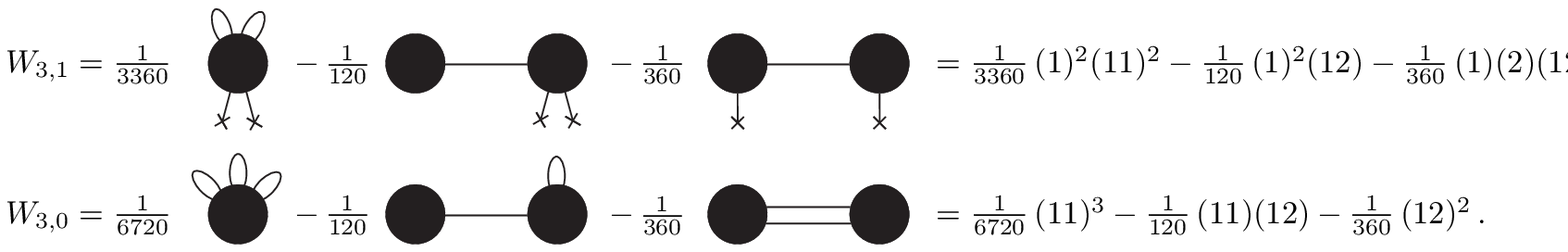}
\end{flushleft}
\end{widetext}
Here we have also introduced
a simple topological notation for the diagrams. The translation between both notations is obvious: $(1)$
is an external source on vertex one, $(11)$ a loop on that vertex, $(12)$ is a link between vertices one and two.
The topological notation makes it possible to present the effective action terms up to level $p=6$ in a relatively compact form:
\begin{widetext}
\begin{eqnarray*}
W_{4,4}&=&\frac{1}{362880}\,(1)^8\, ,\\
W_{4,3}&=&\frac{1}{181440}\,(1)^6 (11)\, ,\\
W_{4,2}&=&\frac{1}{120960}\,(1)^4 (11)^2-\frac{1}{3360}\,(1)^4 (12)-\frac{1}{2520}\,(1)^3 (2) (12)-\frac{1}{2016}\,(1)^2 (2)^2 (12)\\
W_{4,1}&=&\frac{1}{120960}\, (1)^2 (11)^3- \frac{1}{1680}\,(1)^2 (11) (12)-\frac{1}{2520}\,(1)^2 (12)^2-\frac{1}{1260}\, (1)^2 (22)
(12)\\
&& -\frac{1}{2520}\,(1) (2) (11) (12)-\frac{1}{5040}\,(1) (2) (12)^2\, ,\\
W_{4,0}&=&\frac{1}{241920}\,(11)^4 -\frac{1}{2240}\, (11)^2 (12)-\frac{1}{1680}\, (11) (12)^2-\frac{17}{40320}\,(11) (22) (12)\\
&& -\frac{1}{6720}\,(12)^3 +\frac{1}{240}\,(12)(13)\, ,\\
W_{5,5}&=&\frac{1}{39916800}\,(1)^{10}\, ,\\
W_{5,4}&=&\frac{1}{15966720}\,(1)^8 (11)\, ,\\
W_{5,3}&=&\frac{1}{7983360}\,(1)^6 (11)^2-\frac{1}{181440}\,(1)^6 (12)-\frac{1}{90720}\,(1)^5 (2) (12)\\
&& -\frac{1}{25920}\,(1)^4 (2)^2 (12)-\frac{1}{64800}\,(1)^3 (2)^3 (12)\, ,\\
W_{5,2}&=&\frac{1}{5322240}\,(1)^4 (11)^3-\frac{1}{60480}\,(1)^4 (11) (12)-\frac{1}{36288}\,(1)^4 (22) (12)\\
&& -\frac{1}{90720}\,(1)^4 (12)^2-\frac{1}{45360}\,(1)^3 (2) (11) (12)-\frac{1}{45360}\,(1)^3 (2) (12)^2\\
&& -\frac{1}{37800}\,(1)^3 (2) (22)(12)-\frac{1}{15120}\,(1)^2 (2)^2 (11) (12)-\frac{1}{50400}\,(1)^2 (2)^2 (12)^2\, ,\\
W_{5,1}&=&\frac{1}{5322240}\, (1)^2 (11)^4-\frac{1}{40320}\,(1)^2 (11)^2 (12) -\frac{1}{30240}\, (1)^2 (11) (12)^2\\
&& -\frac{1}{17280}\, (1)^2 (11) (22) (12) -\frac{1}{24192}\, (1)^2 (22)^2 (12)-\frac{1}{25200}\, (1)^2 (22) (12)^2\\
&& -\frac{1}{60480}\,(1)^2 (12)^3-\frac{1}{60480}\,(1)(2) (11)^2 (12)-\frac{1}{30240}\, (1) (2) (11) (12)^2\\
&& -\frac{1}{67200}\,(1)(2)(11)(22)(12)-\frac{1}{100800}\,(1) (2) (12)^3+\frac{1}{3360}\, (1)^2 (12) (13)\\
&& +\frac{1}{1260}\, (1)^2 (12) (23) +\frac{1}{2520}\,(1)(2) (12) (13)+\frac{1}{15120}\, (1) (2) (13) (23)\, ,
\end{eqnarray*}
\begin{eqnarray*}
W_{5,0}&=&\frac{1}{10644480}\,(11)^5 -\frac{1}{60480}\,(11)^3 (12)-\frac{1}{30240}\,(11)^2 (12)^2 -\frac{1}{30240}\,(11) (12)^3\\
&& -\frac{1}{151200}\,(12)^4-\frac{13}{403200}\,(11) (22) (12)^2-\frac{11}{241920}\,(11)^2 (22) (12)\\
&& +\frac{1}{2240}\,(11) (12) (13)+\frac{17 }{20160}\,(11) (12) (23)+\frac{1}{1680}\,(12)^2 (13)+\frac{1}{5670}\,(12)(23)(13)\, .
\end{eqnarray*}
\vspace*{0.3cm}
\end{widetext}
Higher-level expressions are more cumbersome and may be found on our web site \cite{speedup}.
Note that the diagrammatic notation reveals the fact that, as far as the short-time
expansion is concerned, all systems fall into
one of two classes of complexity depending on the value of the product $dM$. Discrete-time effective potentials for
systems with $dM>1$ grow faster in complexity with increasing $p$ than their $dM=1$ analogues,
as illustrated in Fig.~\ref{terms}. The reason for this is
that several distinct diagrams collapse into a single term in the case of one particle in one dimension. Symbolical
algebraic calculations for $dM>1$ effective actions were done using the program \cite{speedup} written in Mathematica 6.0
in conjunction with the MathTensor package \cite{mathtensor}. Using this program, for a general many-body theory we have
derived analytic expressions for effective actions up to level $p=10$.

As in the case of one-dimensional systems, the obtained discretized effective actions can be applied to a host of
relevant physical many-body problems in the framework of the Path Integral Monte Carlo approach, including the continuous-space
worm algorithm \cite{boninsegni}. For instance, our approach is applicable to efficiently determine the statistical properties
of Bose-Einstein condensates which are confined in harmonic or anharmonic traps \cite{dalibard2004,dalibard2005,pelster2007,progress}.
Furthermore, our method is ideally suited for dealing with dilute quantum gases in a disorder environment where
the impact of two-particle interactions upon the recently discovered
phenomenon of Anderson localization is at present studied \cite{anderson1,anderson2}.
\section{Conclusions}
We have given a detailed presentation of an analytic procedure for determining the short-time propagation of a
general non-relativistic $M$-particle theory in $d$ dimensions to extremely high orders. The procedure is based
on recursively
solving the Schr\" odinger equation for the transition amplitude in a power series of the propagation time. This
leads to a new recursion relation that has been solved to order $p=10$ for the case of a general
many-body theory. For a single particle moving in a general potential in $d=1$ we have even achieved
$p=35$. In addition, for specific classes of potentials as, for instance, polynomial potentials,
the equations have been
solved to order $p=140$. The resulting Schwinger-Dyson equation and its recursive solution have also
been cast both in a familiar diagrammatic and a compact topological notation. The presented results define
the state-of-the-art for calculating
short-time expansion amplitudes. They can be used to obtain orders of magnitude speedup in Path Integral Monte Carlo
calculations. Thus, the extremely high orders of the short-time expansion make it
possible to perform new and precise analytical
calculations of thermodynamical and dynamical properties of many-body systems.
A list of applications of the presented method to relevant physical systems is briefly outlined.
\section*{Acknowledgements}
We thank Barry Bradlyn for carefully reading the manuscript.
This work was supported in part by the Ministry of Science of the Republic of Serbia, under project No. OI141035,
and the European Commission under EU Centre of Excellence grant CX-CMCS. Symbolical algebraic calculations were
run on the AEGIS e-Infrastructure, supported in part by FP7 projects EGEE-III and SEE-GRID-SCI.
%

\end{document}